# Security Analysis of A Chaos-based Image Encryption Algorithm

Shiguo Lian[*], Jinsheng Sun, Zhiquan Wang

*Department of Automation, Nanjing University of Science and Technology*

*Nanjing, Jiangsu 210094, P.R China*

**Abstract**

The security of Fridrich's algorithm against brute-force attack, statistical attack, known-plaintext attack and select-plaintext attack is analyzed by investigating the properties of the involved chaotic maps and diffusion functions. Based on the given analyses, some means are proposed to strengthen the overall performance of the focused cryptosystem.

## 1. Introduction

With the properties of sensitivity to initial conditions and control parameters, pseudo-randomness and ergodicity, chaotic maps have been widely used in data encryption recently [1-8]. Compared with traditional cryptosystems [9], the ones based on chaos are easier to be realized, which makes it more suitable for large-scale data encryption such as images, videos or audio data. However, there is still not a model for security analysis that is suitable for most chaos-based cryptosystems, which makes that chaotic cryptosystems are seldom accompanied by detailed security analysis [10-16].

A chaos-based image encryption scheme was recently proposed in [17, 18] and is widely referenced afterwards, but its security has not been analyzed efficiently. The encryption scheme is composed of two steps: chaotic confusion and pixel diffusion, where the former process permutes a plain-image with a 2D chaotic map, and the latter process changes the value of each pixel one by one. In the confusion process, the parameters of the chaotic map serve as the confusion key; in the

---

diffusion process, such parameters as the initial value or control parameter of the diffusion function serve as the diffusion key. As shown in Fig. 1, the confusion and diffusion processes are both repeated for several times to enhance the security of this cryptosystem.

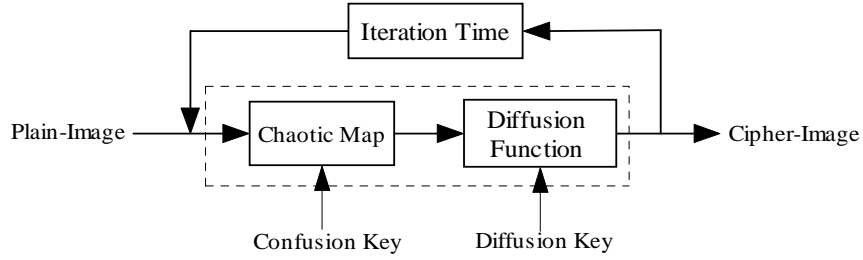

Fig. 1. The chaos-based image encryption scheme proposed in [17, 18]. Here, the plain-image is encrypted into cipher-image through chaotic confusion, diffusion, and repeated iteration.

In the confusion process, many different 2D chaotic maps can be used, such as the Baker map, the Cat map and the Standard, which must be discretized over the image lattice to realize the confusion of all pixels. For a $N \times N$ image lattice, the discretized Standard map, Cat map and Baker map are as follows.

$$\begin{cases} x_{j+1} = (x_j + y_j) \bmod N, \\ y_{j+1} = (y_j + k \sin \dfrac{x_{j+1} N}{2\pi}) \bmod N. \end{cases} \tag{1}$$

$$\begin{bmatrix} x_{j+1} \\ y_{j+1} \end{bmatrix} = \begin{bmatrix} 1 & u \\ v & uv+1 \end{bmatrix} \begin{bmatrix} x_j \\ y_j \end{bmatrix} (\bmod N). \tag{2}$$

$$\begin{cases} x_{j+1} = \dfrac{N}{k_i}(x_j - N_i) + y_j \bmod \dfrac{N}{k_i}, \\ y_{j+1} = \dfrac{k_i}{N}(y_j - y_j \bmod \dfrac{N}{k_i}) + N_i. \end{cases} \text{with} \begin{cases} k_1 + k_2 + \cdots + k_t = N, \\ N_i = k_1 + \cdots + k_{i-1}, \\ N_i \le x_j < N_i + k_i, \\ 0 \le y_j < N. \end{cases} \tag{3}$$

Here, u and v are parameters, ($x_j$, $y_j$) and ($x_{j+1}$, $y_{j+1}$) are the j-th and the j+1-th states, respectively. In Standard map, the confusion key is composed of the parameter *k*. The key is composed of parameters *u* and *v* in Cat map. In Baker map, the key is K=[$k_1$, $k_2$, …, $k_t$] that satisfies the condition proposed in (3).

Fridrich gave some proposals to select suitable diffusion function, according to which, we take two simple ones for example. One is based on addition operation, which is defined as

$$Q_i = (P_i + Q_{i-1}) \bmod L. \tag{4}$$

Here, $P_i$ and $Q_i$ are the i-th plaintext pixel and the i-th ciphertext pixel respectively, L is the pixels' gray-level, $Q_{-1}$ is the initial-value of diffusion process, and is used as diffusion key. The other one is based on power operation, which is defined as

$$Q_i = (P_i + Q_{i-1}^2) \bmod L. \tag{5}$$

Here, the parameters are the same as the ones defined in (4). Fridrich pointed out that the cryptosystem is similar to traditional block ciphers designed following the confusion and diffusion requirements, such as DES and IDEA, but she did not give a detailed security analysis. For example, what kind of chaotic map is better, what kind of diffusion function is more suitable, how to determine the repeated times, and so on.

In this paper, we study the cryptosystem in detail, and give the relation between the involved chaotic map and the security of the cryptosystem, the relation between the diffusion function and the security, and the relation between the repeated times and the overall performance (the security and the complexity). Thus, the chaotic map, the diffusion function and the repeated times can be selected easily according to the revealed relations. What's more, following the revealed relations, some means are proposed to strengthen the overall performance of this cryptosystem.

The rest of the paper is arranged as follows. In Section 2, the functions of the chaotic map and the diffusion function are analyzed, and thus the relation between them and the security of the cryptosystem is constructed. Then, the cryptosystem's security against brute-force attack, statistical attack, known-plaintext attack and select-plaintext attack are analyzed in Section 3. In Section 4, some means are proposed to enhance the security of the cryptosystem. Finally, some conclusions are drawn and future work is presented in Section 5.

## 2. Security Description

In the proposed cryptosystem, 2D chaotic map is used to realize confusion process, and diffusion function is used to realize diffusion process. Thus, the cryptosystem can be regarded as a block cipher similar to traditional block ciphers based on confusion and diffusion operations such as DES, IDEA or NSSU. For convenience of analysis, the cryptosystem is defined as

$$Y = [D(C(X, K_1), K_2)]^n, \qquad (6)$$

where X is plaintext, Y is ciphertext, $K_1$ is confusion key, $K_2$ is diffusion key, and n is iteration time. C() and D() mean confusion process and diffusion process respectively. As can be seen, the cryptosystem's security is determined by the used chaotic map C(), diffusion function D() and iteration time n. Here, the relationship between cryptosystem's security and the used parameters is shown in Table 1.

Table 1. Relationship between the control parameters' properties and the cryptosystem's security.

| Key Space | Key Sensitivity | Plaintext Sensitivity |
|---|---|---|
| Parameter space of chaotic map, Initial-value or parameter space of diffusion function | Parameter sensitivity of chaotic map, Initial-value sensitivity of diffusion function, Iteration time | Initial-value sensitivity of chaotic map, Ergodicity of chaotic map, Spreading speed of diffusion function, Iteration time |

Here, chaotic map's properties are in close relation with the cryptosystem's security. At first, its parameter is used as confusion key. Thus, parameter sensitivity is in close relation with key sensitivity. The higher the parameter sensitivity is, the higher the key sensitivity is, and the stronger the cryptosystem is. Thus, the chaotic map with high parameter sensitivity is preferred in this cryptosystem. Secondly, the initial-value sensitivity and state ergodicity of the chaotic map determine the confusion strength. In chaotic confusion process, initial value refers to the initial position of a pixel. Thus, the higher the initial-value sensitivity is, the smaller the correlation between adjacent pixels is, and the more random the confused image is. Similarly, state ergodicity means that a pixel in certain position can be permuted to any position with the same probability. Thus, the higher the state ergodicity is, the more random the confusion process is, and the more difficult the statistic attack is. Therefore, the chaotic map with high initial-value sensitivity and state ergodicity is preferred in this cryptosystem.

For diffusion function, a change of a pixel can spread to other pixels, which keeps the cryptosystem of high plaintext-sensitivity. The more the pixels are changed in one round of the diffusion process, the higher the diffusion speed of the diffusion function will be. Considering that different diffusion function has different diffusion speed, the one with high diffusion speed is preferred in this cryptosystem.

Besides chaotic map and diffusion function, iteration time is in close relation with cryptosystem's security. The more the iteration time is, the larger the cryptosystem's key space is if different keys are used in different iteration. Also, the more the iteration time is, the higher the cryptosystem's encryption strength is. So bigger n is recommended here in order to obtain high security. Although Fridrich has never proposed to use different keys in different iterations, we will show that it can obtain high security in the following content.

## 3. Security Analysis

### 3.1 Key Space

In the proposed cryptosystem, the confusion process and diffusion process are applied respectively. Thus, the key space of the cryptosystem is the multiplication between the ones of the two processes. Supposing the one of confusion process is $S_1$, and the one of diffusion process is $S_2$, then the one of the cryptosystem is

$$S = S_1 \cdot S_2. \tag{7}$$

Here, the same key is used in different iterations. In practice, different keys can be used in different iterations. If n is the iteration time, and different keys are used in different iterations, then the key space is $S = (S_1 \cdot S_2)^n$. According to Eq. (1)-(5), the confusion space is determined by the parameter space of chaotic map, and the diffusion space is determined by the initial-value space of diffusion function. As can be seen, the cryptosystem's key space S increases with the rise of parameter space $S_1$, initial-value space $S_2$, or iteration time n. Among them, initial-value space is determined by the gray-level of image pixels, parameter space can be adjusted by selecting a suitable chaotic map, and the iteration time can be chosen according to security and complexity requirements. Taking N×N-sized image for example, if the gray-level is L, then the key spaces of the cryptosystems based on the above three chaotic maps are shown in Table 2. Seen from the table, for certain chaotic map, the key space is larger if different key is used in different iteration. For different chaotic maps, it is easy to verify that Standard map has the largest parameter space, Cat map has the smallest one, and the one of Baker map is in the middle. From this view, Standard map and Baker map are preferred than Cat map. Additionally, if the chaotic map is certain, then bigger iteration time is preferred.

Table 2. Comparison of the key space.

| Chaotic Map | Parameter Space | Key Space (the same key in different iteration) | Key Space (different key in different iteration) |
|---|---|---|---|
| Cat map | $N^2$ | $N^2 \cdot L$ | $N^{2n} \cdot L^n$ |
| Baker map | $2^{N-1}$ | $2^{N-1} \cdot L$ | $2^{n(N-1)} \cdot L^n$ |
| Standard map | $(N^2)!$ | $(N^2)! \cdot L$ | $[(N^2)!]^n \cdot L^n$ |

**3.2 Key Sensitivity**

High key sensitivity is required by secure cryptosystems, which means that the ciphertext cannot be decrypted correctly although there is only a slight difference between encryption or decryption keys. This guarantees the security of a cryptosystem against brute-force attacks to some extent. In the proposed cryptosystem, the key sensitivity is determined by the parameter sensitivity of chaotic map and the initial-value sensitivity of diffusion function. And the higher the parameter sensitivity of chaotic map and the initial-value sensitivity of diffusion function are, the higher the cryptosystem's key sensitivity is. Here, the ciphertext change rate caused by key-difference is defined as

$$\frac{\Delta Y}{Y} = \frac{\left|[D(C(X,K_1),K_2)]^n - [D(C(X,K_1+\Delta K),K_2+\Delta K)]^n\right|}{Y}, \quad (8)$$

where $\triangle K$ means a slight difference in key K. As can be seen that, the key sensitivity is in relation with both chaotic map D() and iteration time n.

For 2D chaotic map, we propose to test its parameter sensitivity according to the following method, which is induced from (8). At first, the ciphertexts under different encryption keys (K-|△k|, K and K+|△k|) are computed as below.

$$\begin{cases} Y = [C(X,K)]^n \\ Y_1 = [C(X,K-|\Delta K|)]^n \\ Y_2 = [C(X,K+|\Delta K|)]^n \end{cases}. \quad (9)$$

Here, $[C()]^n$ means to iterate chaotic map for n times, and one bit's difference is applied to encryption key K. Then, the ciphertext difference rate (*Cdr*) is computed as

$$Cdr = \frac{Diff(Y,Y_1) + Diff(Y,Y_2)}{2N^2} \times 100\%, \quad (10)$$

where N is the size of the confused image, and Diff() is a function used to compute the number of different ciphertext-pixels. Here, Diff() is defined as

$$Diff(A,B) = \sum_{i=0}^{N-1}\sum_{j=0}^{N-1} Difp(A(i,j),B(i,j)), \quad (11)$$

where Difp() is

$$Difp(A(i,j),B(i,j)) = \begin{cases} 1, & A(i,j) = B(i,j) \\ 0, & A(i,j) \neq B(i,j) \end{cases}. \quad (12)$$

Here, A and B are two images of the same size, and A(i,j) means the pixel in the i-th row and j-th column in image A. Seen from (9) and (10), *Cdr* is in relation with n, which means that key sensitivity is in relation with iteration time.

For the proposed three chaotic maps, we test their parameter sensitivity. Here, the image size is 256×256. For Standard map, the key is K=*k*, and the key-difference is |△K|=1. For Cat map, the key is K=[u, v], and the key-difference is |△K|=[0, 1] or |△K|=[1, 0]. For Baker map, the key is K=[$k_1, k_2, ..., k_t$], and the key-difference is |△K|=[0, …, 1, –1, 0, …, 0] where "1" should be adjacent to "-1", but the position of them is arbitrary. The relation between key sensitivity and iteration time is shown in Fig. 2, where the iteration time ranges from 1 to 6.

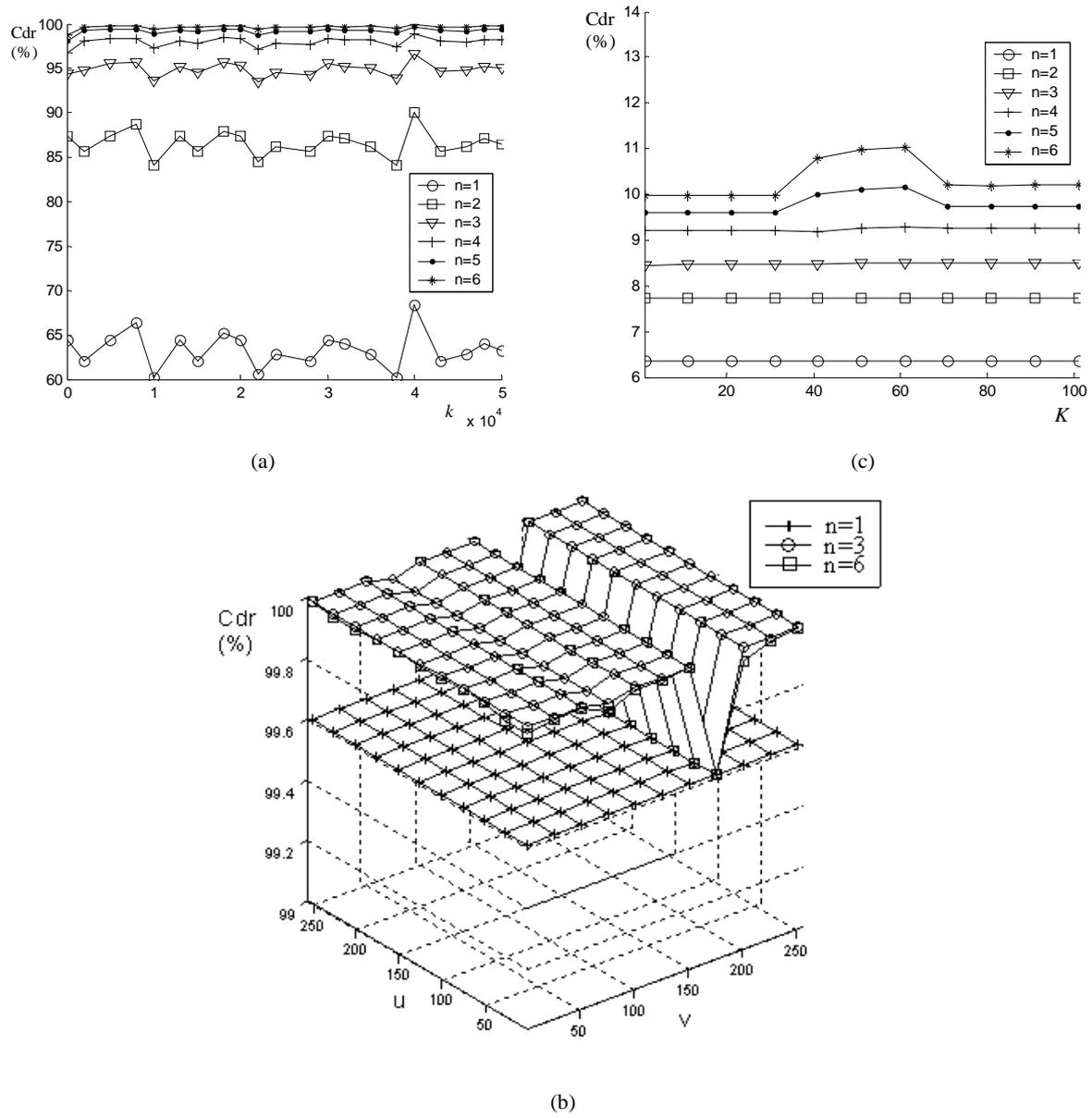

Fig. 2. Key sensitivity test of the three chaotic maps. Here, Fig. (a), (b) and (c) correspond to Standard map, Cat map and Baker map, respectively. In each figure, the curves show the relationship between confusion key, iteration time and ciphertext difference rate (Cdr).

The key sensitivity of Standard map is shown in Fig. 2(a). Where, the key ranges from 0 to 50000. As can be seen, for certain iteration time, the parameters have similar sensitivity; for each parameter, the sensitivity increases with the rise of iteration time. It shows that chaotic map cannot always keep high parameter sensitivity, and big value should be selected for n in order to obtain high key sensitivity. Thus, if Standard map is selected to realize confusion process, the iteration time should be no smaller than 4, which can keep *Cdr* higher than 95%, and thus keep the system of high security.

The key sensitivity of Cat map is shown in Fig. 2(b). Where, the parameters range from 0 to 255. Seen from the figure, the sensitivity keeps always over 99% whenever the iteration times or parameters are. Thus, all the parameters can be used as key even if the iteration time is small. It shows that Cat map is of high parameter sensitivity. This property makes Cat map suitable for data encryption.

The key sensitivity of Baker map is shown in Fig. 2(c). Where, taking the key composed of {56, 2, …, 2} for example, we obtain the keys: $K_0$=[56, 2, …, 2], $K_1$=[2, 56, 2, …, 2], $K_2$=[2, 2, 56, 2, …, 2], …, and $K_{100}$=[2, …, 2, 56]. As can be seen, the test results are similar to the ones on Standard map. That is, for certain parameter, the sensitivity increases with the rise of iteration time, and the sensitivity of different parameters is similar in certain iteration time. The main difference is that the *Cdr* of Baker map is much smaller than the one of Standard map when the iteration time is the same. For example, when n=6, the *Cdr* of Baker map is between 10% and 12%, while the one of Standard map is nearly 100%. The similar result is concluded when comparing with Cat map. It shows that Baker map is of lower parameter sensitivity than Standard map or Cat map. And if it is used in data confusion, the iteration time should be much bigger. Through experiments, we get that the iteration time should be bigger than 12 under the proposed keys in order to keep *Cdr* bigger than 95%.

According to the above analyses and tests, Cat map has the smallest parameter space, but the highest parameter sensitivity. This property makes it suitable for the encryption mode in which different key is used in different iteration. Thus, not only the key space can be enlarged, but also the key sensitivity can be kept in high levels. For Standard map and Baker map, their key spaces are both larger than the one of Cat map. However, the iteration time should be bigger (bigger than 4 for Standard map, and bigger than 12 or above for Baker map) in order to keep high key sensitivity. This property makes them more suitable for the encryption mode in which the same key is used in different iterations.

**3.3 Statistical Attack**

In this cryptosystem, the chaotic map's confusion property is in close relation with the security against statistical attack. If this chaotic map can confuse images to the one with random distribution, then it is difficult for statistical attackers. Here, the confusion strength depends on the chaotic map's properties: initial-value sensitivity and ergodicity. In confusion process, the positions of pixels act as the initial-value of chaotic map. Thus, the higher the chaotic map's initial-value sensitivity is, the less the correlation between adjacent pixels is, and the less the information used by statistical attackers is. On the other hand, the ergodicity of 2D chaotic map shows as the randomness of each pixel. The higher the ergodicity is, the more random each pixel's distribution is, that is to say, each pixel can be permuted to any other pixel with nearly the same possibility. Therefore, the higher the ergodicity is, the more difficult statistical attack is. Thus, the chaotic map with high initial-value sensitivity or high ergodicity is preferred.

High initial-value sensitivity keeps the confusion process of high randomness, which can be concerned by the Lyaponov exponent. However, the discretization process decreases the chaotic maps' properties. What's more, the confusion strength is in relation with both chaotic map C() and iteration time n, which has not been analyzed thoroughly. As have been proposed that, the initial-value sensitivity of 2D chaotic maps acts as position sensitivity of data confusion. Thus, some measurement should be taken to analyze their confusion property. Here, we propose a method to test the confusion property of a chaotic map. That is to test the average distance change (*Adc*) among adjacent pixels. If an N×N image is permuted by chaotic map, the original image is named X, and the permuted one is Y. Thus, the four adjacent pixels (i, j), (i, j+1), (i+1, j) and (i+1, j+1) (i, j=0, 1, …, N-2) in X are permuted to the pixels c(i, j), c(i, j+1), c(i+1, j) and c(i+1, j+1) in Y, respectively. Here, c() is the chaotic map, and c(i, j) is the pixel permuted from the one (i, j). For the four adjacent pixels, the average distance change is defined as

$$Adc(i,j) = (Dis(c(i,j),c(i,j+1)) + Dis(c(i,j),c(i+1,j)) + Dis(c(i,j+1),c(i+1,j+1)) + Dis(c(i+1,j),c(i+1,j+1)))/4 \tag{13}$$

where Dis() is used to compute the distance between two pixels. Here, the confused pixel is defined as

$$(i', j') = c(i, j).$$

And the distance is defined as

$$Dis((i,j),(i',j')) = \sqrt{(i-i')^2 + (j-j')^2}. \tag{14}$$

Thus, the average distance change in the whole image is

$$Adc = \frac{1}{(N-1)^2} \sum_{i=0}^{N-2} \sum_{j=0}^{N-2} Adc(i,j) \times 100\%. \tag{15}$$

Seen from Eq. (14), the average distance change is always bigger than 0, unless the permuted image is the same to the original one. And the bigger *Adc* is, the more confused the original image is. According to the property of chaotic map, the confusion property is in relation with iteration time. Here, *Adc* is in relation with n.

For the proposed three chaotic maps, we test their confusion properties by computing the average distance change. Considering that the confusion property is in relation with chaotic map, iteration time and confusion key, we give the relationship in Fig. 3. Where, the image's size is 256×256, Fig. 3(a), (b) and (c) are the results on Standard map, Cat map and Baker map respectively.

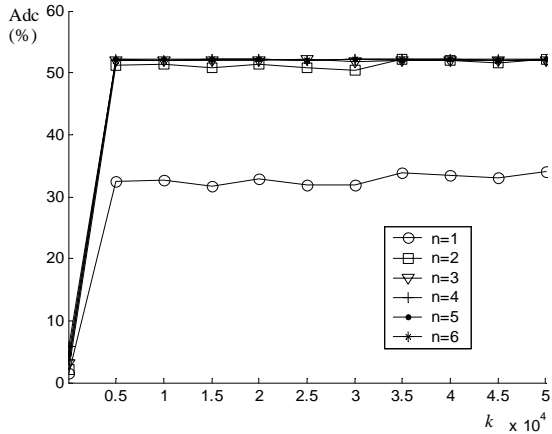

(a)

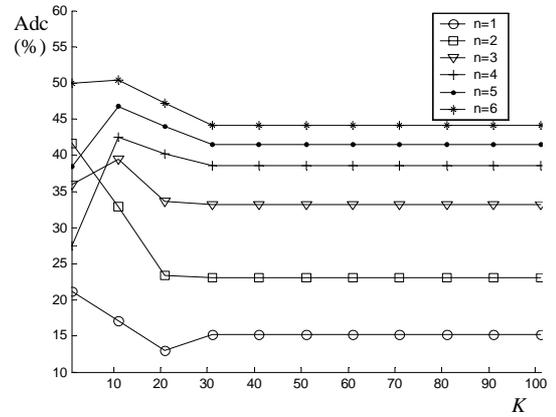

(c)

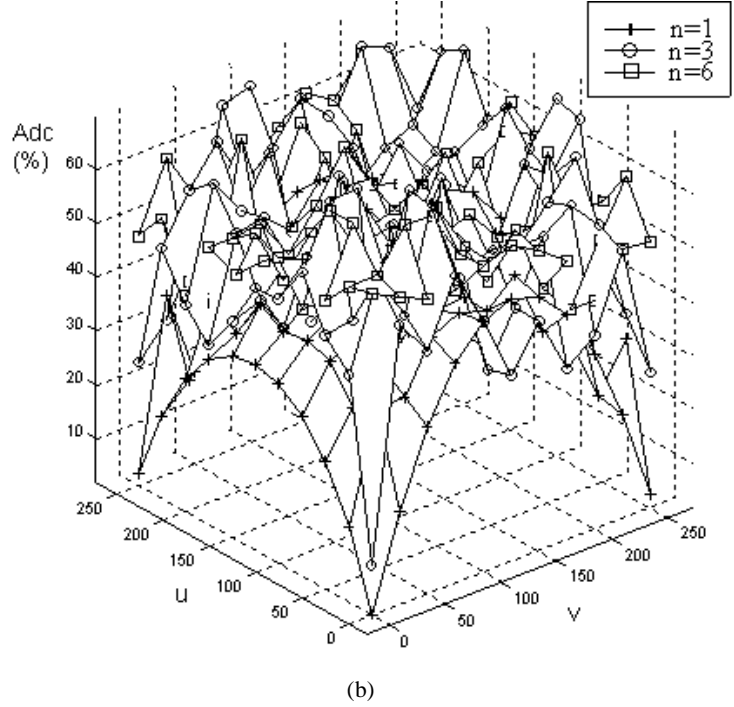

(b)

Fig. 3. Test of average distance change (Adc) of the three chaotic maps. Here, Fig. (a), (b) and (c) correspond to Standard map, Cat map and Baker map, respectively. In each figure, the curves show the relationship between confusion key, iteration time and Adc.

In Fig. 3(a), the chaotic parameter changes from 0 to 50000, and the iteration time ranges from 1 to 6. As can be seen, for different iteration time, the bigger n is, the bigger the average distance change is, and the more random the image is confused; for certain iteration time, the average distance change increases with the rise of parameter $k$ when $k$ is not bigger than 5000, and keeps nearly unchanged when $k$ is bigger than 5000. Generally, it is regarded secure when *Adc* is no smaller than 40%. Under this condition, each pixel-double is estimated in the area making up 64 percents of the whole image, which is difficult for statistic attackers to break. Thus, for Standard map, the parameter should be no smaller than 5000, and the iteration time should be bigger than 1 in order to keep high security against statistic attack.

In Fig. 3(b), the chaotic parameters u and v change both from 0 to 255, and the iteration time ranges from 1 to 6. The results are shown as planes with different colors. For different iteration time, the bigger n is, the higher the according plane is, that is, the more random the image is permuted. For each iteration time, the according plane is nearly plat when n is bigger than 3, which means that different chaotic parameters have the similar confusion property. In order to keep good confusion property (*Adc ≥ 40%*), the iteration time should be no smaller than 6.

In Fig. 3(c), the result is similar with the ones shown in Fig. 3(a) and (b). Where, the confusion parameters are the same to the ones used in Fig. 2(c), which are based on [56, 2, 2, …, 2]. For example, the first confusion parameter is [2, 56, 2, 2, …, 2], the second one is [2, 2, 56, 2, 2, …, 2], …, and the 100-th is [2, 2, …, 2, 56]. As can be seen from Fig. 3(c), the bigger n is, the higher *Adc* is, and the more random the image is permuted. For the same iteration time n, different parameters have the similar confusion property. For this kind of parameters, the iteration time should be bigger than 4 in order to keep good confusion property. Considering of some other keys, the iteration time should be no smaller than 5.

Seen from the test results, for all the three chaotic maps, the confusion property improves with the rise of iteration time. For Cat map, in order to keep high *Adc*, the iteration time should be no smaller than 6, and for Baker map, the iteration time should be bigger than 4. For Standard map, the parameters no smaller than 5000 are preferred, and the iteration time should be bigger than 1. All these conditions keep the chaotic map of good confusion property (*Adc ≥ 40%*), which keeps the confusion process secure against statistic attack.

**3.4 Known-plaintext Attack**

Fridrich has analyzed the security against known-plaintext attack that is to find the similar keys. The difference between the ciphertexts encrypted by different keys is large enough to keep high security against this kind of known-plaintext attack. Note that, there is another known-plaintext attack method, against which, the proposed cryptosystem is not secure enough. In this cryptosystem, the pixels in the corners of the image square have some certain properties. For example, if the chaotic map is Cat map, the pixel of position (0,0) keeps unchanged after any number of iterations. That's to say, if $(x_0, y_0)=(0,0)$, then $(x_0^n, y_0^n)=(0,0)$ after n numbers of chaotic iterations. If the chaotic map is Baker map, the pixel of position (0,0) also keeps unchanged after any number of iterations. That is also testified by the formula of Baker map proposed in [17, 18]. Similarly, in Standard map, the pixel of position (0,0) keeps unchanged after any number of iterations.

As can be seen, (0,0) is the first pixel's position in normal scan mode, which cannot be permuted by chaotic maps. In diffusion process shown in Eq. (4), the fist pixel is always changed by addition operation with diffusion key $Q_{-1}$ during each of the iterations. If the diffusion keys in all the iterations are the same to each other, then it is easy to decrypt diffusion key $Q_{-1}$ even by brute-force attack under the condition of knowing $P_0$. The relationship between the first pixel's plaintext $P_0$, diffusion key $Q_{-1}$ and ciphertext $Q_0$ is

$$Q_0^n = [D(P_0, Q_{-1})]^n, \qquad (16)$$

where $Q_0^n$ is the fist pixel's ciphertext after n iterations, D() means the diffusion function shown in Eq. (4) or (5), and $[D()]^n$ means to apply diffusion process for n times. Taking diffusion function (4) for example, Eq. (16) can be induced as

$$\begin{aligned} Q_0^n &= (Q_0^{n-1} + Q_{-1}) \bmod L = [(Q_0^{n-2} + Q_{-1}) \bmod L + Q_{-1}] \bmod L \\ &= (Q_0^{n-2} + 2Q_{-1}) \bmod L = \cdots = (Q_0^0 + nQ_{-1}) \bmod L \\ &= (P_0 + nQ_{-1}) \bmod L \end{aligned} \qquad (17)$$

Here, plaintext $P_0$, ciphertext $Q_0^n$, iteration time n, pixel gray-level L and diffusion function D() are all known to attackers, so it is easy to decode the diffusion key according to the following process.

$$nQ_{-1} \bmod L = (Q_0^n - P_0) \bmod L \Rightarrow \begin{cases} Q_{-1} = \dfrac{kL + (Q_0^n - P_0) \bmod L}{n}, & Q_{-1} \in Z \text{ and } Q_{-1} \in [0, L-1] \\ 0 \le k \le n-1, & k \in Z \end{cases}. \qquad (18)$$

Similarly, for diffusion function (5), the decoding process is

$$nQ_{-1}^2 \bmod L = (Q_0^n - P_0) \bmod L \Rightarrow \begin{cases} Q_{-1} = \sqrt{\dfrac{kL + (Q_0^n - P_0) \bmod L}{n}}, & Q_{-1} \in Z \text{ and } Q_{-1} \in [0, L-1] \\ 0 \le k \le n-1, & k \in Z \end{cases}. \qquad (19)$$

As can be seen from (18) and (19), the decoding process is to find $Q_{-1}$ that satisfies two conditions: it is an integer, and it locates in [0, L-1]. Thus, the computing process is composed of n times of attempts, and each attempt includes only few addition/subtraction, multiplication/division and comparison operations. For example, in (18), each attempt includes 3 times of addition/subtraction operations, 2 times of multiplication/division operations and once comparison operation. While in (19), it includes 4 times of addition/subtraction operations, 2 times of multiplication/division operations and once comparison operation. As is known, iteration time n is often selected no bigger than 64 in order to keep high speed. Thus, the cost of attack is so little that it is very easy for known-plaintext attackers to break diffusion key. What's more, if the diffusion key is broken, then the brute-force space of the cryptosystem will decrease L times. Thus, the cryptosystem is vulnerable even to brute-force attacks, without considering of other more powerful attacks. So it is necessary to solve this problem in order to keep high security.

**3.5 Select-plaintext Attack**

In select-plaintext attack, some plaintexts with little difference may be selected to analyze the according difference between their ciphertexts. Thus, much difference between ciphertexts is expected in order to keep high security. This is in relation with key sensitivity and plaintext sensitivity. Among them, key sensitivity has been analyzed, so we emphasize on plaintext sensitivity here. If a change happens in a pixel's gray-level, then the change can cause great ones in other pixels through diffusion process. Thus, the greater the changes caused by diffusion process are, the higher the cryptosystem's plaintext sensitivity is, and the more difficult the system's security against differential attack is. Here, we define the difference caused by plaintext difference as

$$\frac{\Delta Y}{Y} = \frac{\left| [D(X, K_2)]^n - [D(X+\Delta X, K_2)]^n \right|}{Y}, \quad (20)$$

where $\triangle X$ means a slight gray-level difference in plaintext X. As can be seen, the diffusion process is in relation with diffusion function D() and iteration time n.

In order to analyze diffusion properties, we propose a method to test the relationship between diffusion speed and iteration time. The diffusion speed is measured by pixel change rate (*Pcr*) that means the ratio between the number of changed pixels and the image's size when a slight gray-level difference happens. And it is defined as

$$Pcr = \frac{Dif(Y, Y')}{2N^2} \times 100\%. \quad (21)$$

where Dif() has been defined in Eq. (11) and (12), and N is image's size. Y and Y' are defined respectively as

$$\begin{cases} Y = [D(X, K_2)]^n \\ Y' = [D(X+\Delta X, K_2)]^n \end{cases}, \quad (22)$$

where $\triangle X$ means a slight change in one pixel's gray-level, for example, one bit's change. Thus, for each plaintext image, the relationship between pixel change rate and iteration time can be shown by *Pcr-n* curve. From the curves, the suitable diffusion function with high diffusion speed can be selected.

In $256 \times 256$-sized image Lena, we test the diffusion speed of the diffusion functions shown in Eq. (4) and (5). The *Pcr-n* curves are shown in Fig. 4. As can be seen, for each of them, *Pcr* increases with the rise of iteration time n, and reaches 100% when n is about 5. But the curve according to function (4) oscillates between 50% and 100% after *Pcr* has reached 100%, which means that the diffusion process is not stable enough for any kind of image data. While the one according to function (5) keeps about 100% when *Pcr* has reached 100%, which means that this kind of diffusion process is stable for various image data. In order to keep high diffusion speed and to obtain high security against differential attacks, the diffusion function with stable diffusion curve and high diffusion speed is preferred. So we recommend the one shown in function (5) here. Of course, the condition should be satisfied that iteration time n is not smaller than 4.

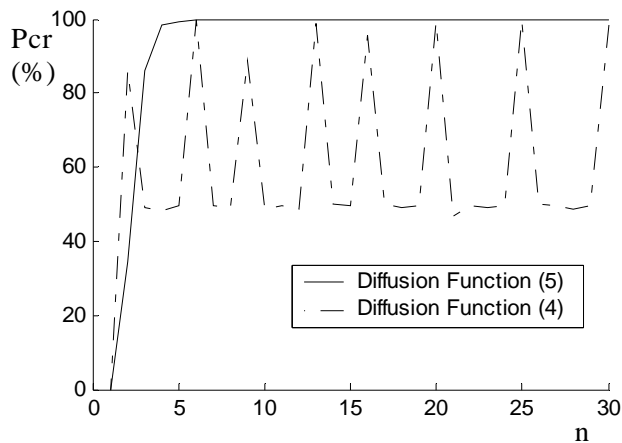

Fig. 4. Diffusion speed test of the two diffusion functions. The solid and broken lines correspond to diffusion function (5) and (4), respectively.

**3.6 Security and Complexity**

The cryptosystem's security is in relation with its computing complexity. And its computing complexity depends on iteration time n, the computing complexity of 2D chaotic map and the one of diffusion function. Among them, the high cost caused by chaotic map or diffusion function can be decreased through selecting suitable one according to its computing complexity. Here, we give the computing complexity of the above three chaotic maps and two diffusion functions, which is shown in Table 3. Where, the statistic is made on an N×N-sized image, the cost of once addition or subtraction operation is represented by a, and the one of multiplication or division operation is b. Seen from the table, each of the diffusion functions or chaotic maps has different computing complexity, which is guidance for designers. For example, Baker map or diffusion function (4) can be selected in order to get high speed.

Table 3. Comparison of computing complexity.

| Operation | Chaotic Map | | | Diffusion Function | |
|---|---|---|---|---|---|
| | Cat map | Baker map | Standard map | Function (4) | Function (5) |
| Addition/Subtraction | $2N^2 a$ | $2N^2 a$ | $2N^2 a$ | $1N^2 a$ | $1N^2 a$ |
| Multiplication/Division | $3N^2 b$ | $2N^2 b$ | $4N^2 b$ | $1N^2 b$ | $2N^2 b$ |

As has been proposed that the n with bigger value is recommended in order to get high security. This includes two cases: 1) the confusion key or diffusion key is unique in n times of iteration; 2) the keys in n iterations are different from each other. In the first case, the confusion mode needs only to be computed once, for the parameter of chaotic map keeps unchanged during all these iterations. While in the second case, a novel confusion mode should be computed for each of these iterations because the parameters are different in these iterations. Thus, taking N×N-sized image for example, if the computing-complexity of once chaotic confusion is $R_c(N)$, the one of once diffusion process is $R_d(N)$, and the iteration time is n, then the cryptosystem's computing-complexity in Case 1 or Case 2 is computed as below respectively.

$$\begin{cases} R_1(N) = R_c(N) + n \cdot R_d(N) \\ R_2(N) = n \cdot [R_c(N) + R_d(N)] \end{cases}. \tag{23}$$

Here, Case 1 is of lower computing complexity than Case 2, which is based on the degradation of security. If the key space of once chaotic diffusion is $K_c(N)$, and the one of diffusion process is $K_d(N)$, then the key spaces of Case 1 and Case 2 are computed as below respectively.

$$\begin{cases} K_1(N) = K_c(N) \cdot K_d(N) \\ K_2(N) = [K_c(N) \cdot K_d(N)]^n \end{cases}. \tag{24}$$

As can be seen that, the key space of Case 1 is smaller than the one of Case 2, but the computing-complexity of Case 1 is lower than the one of Case 2. What's more, according to the property of chaotic map, it should be iterated for some times with the same parameter in order to enhance the confusion strength. Thus, Case 1 is more secure than Case 2 in statistic or differential attack. As can be seen, the security contradicts with the computing complexity. Therefore, a tradeoff between security and computing complexity should be made in order to improve its practical applications.

## 4. Security Improvement

### 4.1 Parameter Selection

According to the above analyses and experiments, we propose the following advice to choose suitable chaotic map, diffusion function and iteration time n:

1) Different chaotic map has different confusion property that is in relation with iteration time. Enough iteration time should be satisfied in order to keep high security against statistic attack or differential attack. Taking the proposed three chaotic maps for example, the iteration time for Standard map should be no smaller than 4, the one for Cat map is no smaller than 6, and the one for Baker map should be no smaller than 12 or above.

2) Among the proposed three chaotic maps, Standard map has the largest parameter space and the highest computing complexity, Baker map has the lowest computing complexity and middle parameter space, and Cat map has the smallest parameter space and the middle computing complexity. Thus, Baker map is preferred as a tradeoff between security and computing complexity.

3) For high security, diffusion function shown in Eq. (5) is preferred, in order to keep the cryptosystem of high plaintext sensitivity and thus secure against differential attacks.

Certainly, some other chaotic maps or diffusion functions can also be used here, if they satisfy the conditions proposed in Section 3, or can get better tradeoff between security and computing complexity.

**4.2 Permuting (0,0)-Pixel**

As has been analyzed in Section 3.5, the pixel in corner (0,0) cannot be permuted to other positions, which is a threat to the whole cryptosystem. In order to avoid it, a method can be adopted to change the position of the corner-pixel. That is, to change the scan order after each chaotic map, which changes the position of the first pixel. Thus, it is difficult to get the ciphertext $Q_0^n$, which increases the difficulty of breaking the diffusion key.

**4.3 Introducing Iteration Time $n_0$**

As has been analyzed in Section 3.6, the cryptosystem's security contradicts with its computing complexity. In order to solve this contradiction, a method can be used to get a suitable tradeoff. That is, to divide n times into $n/n_0$ groups and use different key in different group. In each group, the chaotic map is iterated with the same key for $n_0$ times. Thus, the computing complexity and key space are computed respectively as

$$\begin{cases} R(N) = \dfrac{n}{n_0} R_c(N) + n \cdot R_d(N) \\ K(N) = [K_c(N) \cdot K_d(N)]^{\frac{n}{n_0}} \end{cases}. \quad (25)$$

Thus, the computing complexity and key space of the modified case are between the ones of Case 1 and the ones of Case 2: $R_1(N) \le R(N) \le R_2(N)$ and $K_2(N) \le K(N) \le K_1(N)$.

**4.4 Introducing Key Generator**

In this cryptosystem, the key is composed of two parts: confusion sub-key and diffusion sub-key. In order to keep high security, it is necessary to introduce a key generator to realize sub-key generation and distribution. Thus, for the encryption process with n-time repeating, $n/n_0$ key-troubles are generated in the encryption process that is symmetric with the decryption one. Based on the chaotic logistic map, a key generator is proposed here. The user key K is divided into two parts ($K_1$ and $K_2$) that are used as the initial-value of chaotic logistic map. Let $X_t^i$ (i=1,2 and t=1, 2, 3, …, $\frac{n}{n_0}$) denote the key in the t-th iteration, it is defined as

$$\begin{cases} x_t^1 = f^T(\dfrac{x_{t-1}^1 + x_{t-1}^2}{2}), \\ x_t^2 = f^T(\dfrac{\left| x_{t-1}^1 - x_{t-1}^2 \right|}{2}), \\ X_t^i = \sum_{j=0}^{S-1} x_t^i(j) \cdot 2^j. \end{cases} \quad (26)$$

Here, $x_0^i = K_i$ (i=1, 2), and $S$ is the sub-key's length. $f^T()$ means to iterate the chaotic logistic map for T times, and T=100 is recommended. Thus, a slight change in user key K will cause great changes in the sub-keys, which keeps the cryptosystem of high key sensitivity.

**5. Conclusions and Future Work**

The proposed image encryption system is analyzed thoroughly in this paper. Through studying the strengths of the confusion and diffusion properties, its security against statistical attack, known-plaintext attack, select-plaintext attack and so on, some enhancement means are proposed to improve the cryptosystem, and some advices is given to select suitable chaotic map, diffusion function and iteration time. Of course, some other chaotic maps or diffusion functions can also be used in the studied cryptosystem, if they satisfy the conditions discussed in this paper, or can get a better tradeoff between the security and the computing complexity. That is our future work.

**Acknowledgements**

The authors thank Dr. Shujun Li for useful discussion and are also indebted to the referee for some important suggestions, and thank Dr. Jinwei Wang for his valuable help. This research was supported by the National Natural Science Foundation of China through the grant number 60174005 and the Natural Science Foundation of Jiangsu Province through the grant number BK2004132.